# Mathematical Model of a Direct Methanol Fuel Cell

Brenda L. García [a], Vijay A. Sethuraman[a], John W. Weidner[a,1], Roger Dougal[b], and Ralph E. White[a]

[a]Center for Electrochemical Engineering

Department of Chemical Engineering

University of South Carolina

Columbia, SC 29208, USA

[b]Department of Electrical Engineering

University of South Carolina

Columbia, SC 29208, USA

[1] Corresponding author. Tel.: 1-803-777-3207; Fax: 1-803-777-8265.

*E-mail address:* weidner@engr.sc.edu (J.W. Weidner).

# ABSTRACT

A one dimensional (1-D), isothermal model for a direct methanol fuel cell (DMFC) is presented. This model accounts for the kinetics of the multi-step methanol oxidation reaction at the anode. Diffusion and crossover of methanol are modeled and the mixed potential of the oxygen cathode due to methanol crossover is included. Kinetic and diffusional parameters are estimated by comparing the model to data from a 25 cm$^2$ DMFC. This semi-analytical model can be solved rapidly so that it is suitable for inclusion in real-time system level DMFC simulations.

# INTRODUCTION

Direct Methanol Fuel Cells (DMFCs) are currently being investigated as alternative power source to batteries for portable applications because they can offer higher energy densities. However, two factors limit the performance of DMFC systems: crossover of methanol from anode to cathode and the slow kinetics of the electrochemical oxidation of methanol at the anode.

The crossover of methanol lowers the system efficiency and decreases cell potential due to corrosion at the cathode. Figure 1 illustrates the electrochemistry and transport phenomena in DMFCs. Electrochemical oxidation of methanol occurs at both anode and cathode, but corrosion current at the cathode produces no usable work. Several experimental and modeling studies have characterized methanol crossover in DMFCs [1-4].

The kinetics of DMFCs are complicated because the reaction mechanism involves adsorption of methanol and several reaction steps including the oxidation of CO. Figure 2 shows a possible network of reaction pathways by which the electrochemical oxidation of methanol occurs. Catalysis studies have attempted to analyze possible reaction pathways to find the main pathway of methanol oxidation [6-8]. Most studies conclude that the reaction can proceed according to multiple mechanisms. However, it is widely accepted that the most significant reactions are the adsorption of methanol and the oxidation of CO. Follows is a simplified reaction mechanism that will be used in this paper to model performance of DMFCs.

$$CH_3OH + Site \rightarrow (CH_3OH)_{ads} \qquad (1)$$

$$(CH_3OH)_{ads} \rightarrow (CO)_{ads} + 4H^+ + 4e^- \qquad (2)$$

$$(CO)_{ads} + H_2O \rightarrow CO_2 + 2H^+ + 2e^- \tag{3}$$

This mechanism is similar to the mechanism used by Meyers and Newman [9], but does not segregate the electrochemical oxidation of water reaction from the electrochemical oxidation of CO. This assumption does not change the kinetic expression appreciably and is applicable for Pt-Ru catalysts where the oxidation of water on Ru occurs much faster than the oxidation of CO.

The model presented in this paper seeks to provide a one dimensional (1-D), isothermal model of a DMFC that allows rapid prediction of polarization data and gives insight into mass transport phenomena occurring in the cell. Models currently in the literature leave out effects important for predicting full cell performance or include physical detail that encumbers the model and complicates its solution. Baxter et al.[10] developed a model for the DMFC anode which considers diffusion of $CO_2$, $H_2O$ and methanol in the anode, but neglects the effects of the cathode and thus does not capture the effects of methanol crossover. They also used Butler-Volmer kinetics to describe the electrochemical oxidation of methanol. Meyers and Newman [9] develop a kinetic expression similar to the one used in this paper and provide parameters for the cathode reaction, but do to the level of detail included in their membrane model the non-linearity of their equations make the solution of the model difficult. Kulikovsy [11] solved an analytical model for the fuel cell anode to predict the anodic overpotential. However, the model presented could only be solved in the limits of low current or high current and did not allow prediction of full cell polarization behavior. Wang and Wang [12] used a CFD model to investigate a full DMFC fuel cell. This analysis included two-phase flow effects in the backing layers (BLs) but used a non-intuitive transition in kinetics at a certain concentration to describe the complex methanol oxidation reaction. Norlund and Lindbergh [13] develop an anode model that neglects the effects of methanol crossover and the cathode. Their model also assumes a flooded agglomerate model of the anode catalyst layer (ACL) that assumes a specific geometry for all reaction sites.

## EXPERIMENTAL

*Cell Preparation*

Tests were performed on a 25 $cm^2$ fuel cell from Fuel Cell Technologies. The membrane electrode assembly (MEA) was constructed from a Nafion® 117 membrane with E-TEK 40% Platinum/C gas diffusion electrodes prepared according to the method of Wilson [14]. The anode loading was 3 $mg/cm^2$ of 1:1 Pt/Ru catalyst and the cathode loading was 1 $mg/cm^2$ of Pt. Tests were conducted using an 890C load cell from Scribner Associates Inc. with a methanol fuel system. The cell was broken in by running for 3 hours under a 5 A load with a 40 mL/min flow of 1 M feed to the anode and 50 mL/min flow of dry oxygen on the cathode. The cell temperature and inlet temperatures were 70°C. All reagents were certified as ultra high purity.

*Testing*

Prior to running tests with a given concentration of methanol the system was flushed with 1.5 L of methanol. The flow rates for the anode and cathode were then set to those necessary to maintain 5/5 stoichiometric excess ratios on the anode and cathode. The minimum flow rate for all experiments was 10 mL/min on the anode and 50 mL/min on the cathode. The cell was next run under a load of 0.25A for 10 minutes or until the voltage reached steady-state. The load was set to 0 A for 10 minutes or until the voltage arrived at its steady open circuit value. Polarization curves were run in current scan mode with 0.02 A/point and 150 seconds/point.

# MODEL DEVELOPMENT

*Assumptions*

The model presented here accounts for concentration variations of methanol across the anode backing layer (ABL), anode catalyst layer (ACL), and membrane. Figure 3 presents a schematic diagram of the layers considered in the model illustrating several assumptions. The assumptions used in this model are

1. Steady-state.
2. Variations in only one spatial Cartesian coordinate (i.e., across the MEA).
3. Convective transport of methanol is negligible.
4. Pressure gradient across the layers are negligible.
5. Isothermal conditions.
6. All physical properties are considered constant.
7. Only liquid phase is considered. This means that carbon dioxide remains dissolved in solution.
8. Solutions are considered ideal and diluted.
9. Local equilibrium at interfaces can be described by a partition function.
10. The ACL is assumed to be a macro-homogeneous porous electrode and thus the reaction in this layer is modeled as a homogeneous reaction.
11. Anode kinetics can be described by the step mechanism, Eq. (1) – (3), with a rate expression similar to the one obtained by Meyers and Newman [9].
12. The anodic overpotential is constant throughout the ACL.

13. Cathode kinetics can be described by Tafel expression with no mass transfer limitations.

Applying these assumptions, the mass transport equations are developed and combined with the kinetic equations in order to calculate the cell voltage, which can be expressed as:

$$V_{Cell} = U^{O_2} - U^{MeOH} - \eta_C - \eta_A - \frac{\delta_M I_{Cell}}{\kappa} \qquad (4)$$

where $U^{O_2}$ and $U^{MeOH}$ are the thermodynamic equilibrium potential of oxygen reduction and methanol oxidation respectively, $\eta_C$ and $\eta_A$ are the cathode and anode overpotentials, respectively, and the last term in Eq. (4) represents the ohmic drop across the membrane.

### *Governing Equations and Boundary Conditions-Anode*

The anode overpotential is obtained by first obtaining the concentration profiles across the various regions of the MEA.

<u>Anode Backing Layer</u>

The differential mass balance for methanol in the ABL is

$$\frac{dN^B_{MeOH,z}}{dz} = 0 \qquad (5)$$

Assuming Fickian diffusion [15] of methanol with an effective diffusivity $D_B$ in the ABL phase, the methanol flux gives

$$N^B_{MeOH,z} = -D_B \frac{dc^B_{MeOH}}{dz} \qquad (6)$$

Substitution of Eq. (6) into Eq. (5) gives the governing equation for methanol in the ABL as

$$\frac{d^2 c_{MeOH}^B}{dz^2} = 0 \tag{7}$$

The boundary conditions for Eq. (7) are illustrated in Fig. 3. It is assumed that concentration at the flow-channel/ABL interface is given by the bulk concentration in the flow channel. The concentration at the ABL/ACL interface is given by assuming local equilibrium with a partition coefficient $K_I$.

$$\text{At } z = 0: \quad c_{MeOH}^B = c_b \tag{8}$$

$$\text{At } z = z_I: \quad c_{MeOH}^B = c_I^B = K_I c_I^A \tag{9}$$

Membrane

The differential mass balance for methanol in the membrane is

$$\frac{dN_{MeOH,z}^M}{dz} = 0 \tag{10}$$

The transport of methanol in the membrane is governed by diffusion and electro-osmotic drag. The flux equation can be written as

$$N_{MeOH,z}^M = -D_M \frac{dc_{MeOH}^M}{dz} + \xi_{MeOH} \frac{I_{Cell}}{F} \tag{11}$$

where $D_M$ and $\xi_{MeOH}$ are the effective diffusion and the electro-osmotic drag coefficients of methanol respectively. The electro-osmotic drag coefficient is defined as the number of methanol molecules dragged by a hydrogen ion moving in the membrane. Substitution of Eq. (11) into Eq. (10) gives the governing equation for methanol in the membrane as

$$\frac{d^2 c_{MeOH}^M}{dz^2} = 0 \tag{12}$$

The boundary conditions for Eq. (12) are illustrated in Fig. 3. It is assumed that all the methanol crossing the membrane reacts at the cathode creating a very low concentration at the membrane/cathode-layer interface. The concentration at the ACL/membrane interface is given by assuming local equilibrium with a partition coefficient $K_{II}$.

$$\text{At } z = z_{II}: \quad c_{MeOH}^{M} = c_{II}^{M} = K_{II} c_{II}^{A} \tag{13}$$

$$\text{At } z = z_{III}: \quad c_{MeOH}^{M} \approx 0 \tag{14}$$

Anode Catalyst Layer

The methanol oxidation reaction at the anode is considered homogeneous. The differential mass balance for methanol in the ACL is

$$\frac{dN_{MeOH,z}^{A}}{dz} = \frac{r_{MeOH}}{M_{MeOH}} \tag{15}$$

where the molar consumption rate $(r_{MeOH}/M_{MeOH})$ is related to the volumetric current density $j$ as

$$\frac{r_{MeOH}}{M_{MeOH}} = \frac{-j}{6F} \tag{16}$$

The current density expression for methanol oxidation is taken from Meyers and Newman [9] as

$$j = a I_{0,ref}^{MeOH} \frac{k c_{MeOH}^{A}}{c_{MeOH}^{A} + \lambda e^{\frac{\alpha_{A} \eta_{A} F}{RT}}} e^{\frac{\alpha_{A} \eta_{A} F}{RT}} \tag{17}$$

where $a$ is the specific surface area of the anode, $I_{0,ref}^{MeOH}$ is the exchange current density, and $k$ and $\lambda$ are constants.

The methanol flux in the ACL with an effective diffusivity $D_{A}$ is given by a similar expression as showed for the ABL.

$$N^A_{MeOH,z} = -D_A \frac{dc^A_{MeOH}}{dz} \tag{18}$$

Substitution of Eq. (16) and (18) into Eq. (15) gives the governing equation for methanol in the ACL as

$$D_A \frac{d^2 c^A_{MeOH}}{dz^2} = \frac{j}{6F} \tag{19}$$

The boundary conditions for Eq. (19) are illustrated in Fig. 3. The methanol concentration at the interfaces is given as

$$\text{At } z = z_I: \quad c^A_{MeOH} = c^A_I \tag{20}$$

$$\text{At } z = z_{II}: \quad c^A_{MeOH} = c^A_{II} \tag{21}$$

The concentrations given in Eq. (20) and (21) are related to the concentrations at the ABL and the membrane through Eq. (9) and (13). These concentrations can be determined from jump mass balances [15] at the ABL/ACL and ACL/membrane interfaces, yielding

$$\text{At } z = z_I: \quad N^B_{z,MeOH} = N^A_{z,MeOH} \tag{22}$$

$$\text{At } z = z_{II}: \quad N^A_{z,MeOH} = N^M_{z,MeOH} \tag{23}$$

*Analytical Solution-Anode*

The solution to Eq. (7) – (9) is

$$c^B_{MeOH} = \frac{K_I c^A_I - c_b}{\delta_B} z + c_b \tag{24}$$

The solution to Eq. (12) – (14) is

$$c^M_{MeOH} = K_{II} c^A_{II} \left( \frac{\delta_B + \delta_A - z}{\delta_M} + 1 \right) \tag{25}$$

The solution to Eq. (19) – (21) is:

$$c_{MeOH}^{A} = \frac{I_{Cell}}{12F\delta_A D_A} z^2 + C_1 z + C_2 \quad (26)$$

where

$$C_1 = \frac{c_{II}^{A} - c_{I}^{A}}{\delta_A} - \frac{I_{Cell}(2\delta_B + \delta_A)}{12F\delta_A D_A} \quad (27)$$

and

$$C_2 = c_I^{A} - \frac{(c_{II}^{A} - c_{I}^{A})\delta_B}{\delta_A} + \frac{I_{Cell}\delta_B(\delta_B + \delta_A)}{12F\delta_A D_A} \quad (28)$$

From the solutions above the fluxes in each phase can be obtained via Eq. (6), (11), and (18). The fluxes are then evaluated at the respective interfaces to obtain two expressions in terms of $c_I^A$ and $c_{II}^A$ from Eq. (21) and (22). One may ultimately show that

$$c_I^A = \frac{\delta_A D_M K_{II}\left(D_B c_b - \frac{I_{Cell}\delta_B}{12F}\right) + \delta_M D_A\left(D_B c_b - (1+6\xi_{MeOH})\frac{I_{Cell}\delta_B}{6F}\right)}{D_B K_I(\delta_A D_M K_{II} + \delta_M D_A) + \delta_B D_A D_M K_{II}} \quad (29)$$

$$c_{II}^A = \frac{\delta_M\left(D_A D_B c_b - \delta_A D_B K_I(1+12\xi_{MeOH})\frac{I_{Cell}}{2nF} - \delta_B D_A(1+6\xi_{MeOH})\frac{I_{Cell}}{6F}\right)}{D_B K_I(\delta_A D_M K_{II} + \delta_M D_A) + \delta_B D_A D_M K_{II}} \quad (30)$$

Finally, the concentration profile given by Eq. (26) is substituted into the kinetic expression, Eq. (17), integrated, and equated to the cell current giving

$$I_{Cell} = \int_0^{\delta_A} aI_{0,ref}^{MeOH} \frac{kc_{MeOH}^A}{c_{MeOH}^A + \lambda e^{\frac{\alpha_A \eta_A F}{RT}}} e^{\frac{\alpha_A \eta_A F}{RT}} dx \quad (31)$$

Assuming $\eta_A$ is constant (assumption 12), Eq. (31) is used to obtain $\eta_A$ for a given value of $I_{Cell}$.

*Cathode*

Tafel kinetics with first order oxygen concentration dependence is employed to describe the oxygen reduction at the cathode.

$$I_{Cell} + I_{leak} = I_{0,ref}^{O_2} \frac{c_{O_2}}{c_{O_2,ref}} e^{\frac{\alpha_C \eta_{C,mix} F}{RT}} \tag{32}$$

where $I_{leak}$ is the leakage current density due to the oxidation of methanol crossing the membrane. The leakage current density can be written as

$$I_{leak} = 6FN_{MeOH,z}^{M} \tag{33}$$

where $N_{MeOH,z}^{M}$ is obtained from Eq. (11). Equation (32) is then used to obtain $\eta_C$ for a given value of $I_{Cell}$.

Finally, the anode and cathode overpotentials are substituted into Eq. (4) to give give $V_{Cell}$ for a given value of $I_{Cell}$.

## RESULTS AND DISCUSSION

Experimental and modeling results of polarization behavior for 0.05M, 0.1M, 0.2M, and 0.5M methanol solutions are shown in Fig. 4. The limiting current densities predicted by the model are very close to experimental values. The model predictions for conditions near open circuit voltage show the largest errors with experimental values. This disagreement could be due to the fact that concentration and temperature effects on the thermodynamic potentials of the electrodes were neglected. Methanol polarization data above 0.5M could not be modeled with the same set of kinetic and transport parameters as was used for the cases shown in Fig. 4. Trends in the predicted and modeled polarization curves in Fig. 4 are similar to those shown for 0.2M and 0.5M in Wang and Wang [12]. However, the limiting current densities Wang and Wang [12] predict are higher than those in Fig. 4. In their paper, they contend that high current densities in DMFCs can be explained by the possibility of gas phase transport.

The modeling parameters used are listed in Table 1. Transport parameters agree well with literature values. The specific area ($a$) and the anode and cathode transfer coefficients can change due to electrode properties and were adjusted to fit the model to the experimental data. It was found that around the parameter set listed in Table 1 certain parameters could be adjusted simultaneously and the resulting fit did not alter the polarization curves significantly. One example is that increasing the exchange current density while increasing $\lambda$ produced nearly equivalent curves. For this reason, all parameters in Table 1 are listed only to two significant digits. For the model development the methanol electro-osmotic drag coefficient was assumed to be a constant value, but when solving the model the methanol electro-osmotic drag coefficients was estimated at every point on the polarization curve according to the equation in Table 1.

Figure 5 shows predicted concentration profiles across the anode and membrane for the three concentrations at 15 mA/cm$^2$. At this condition a cell operating with a 0.05M bulk methanol concentration is in the mass transfer limited region while the 0.1M, 0.2M, and 0.5M concentrations are in the region limited by the oxidation of CO on the catalyst surface. The concentration for the 0.05M case in the catalyst layer is very low at this current density similar to what should be expected. The concentration profile across the catalyst layer appears to be nearly constant for the 0.05M, 0.1M, and 0.2M concentrations. The 0.5M concentration has a larger drop in concentration across the catalyst layer due to a higher rate of methanol crossover, but the value is still relatively constant. The assumption that the methanol concentration in the ACL is constant is most valid close to the limiting current density where the methanol concentration is the lowest, thus reducing the amount of methanol crossover.

Figure 6 shows calculations of the methanol crossover predicted by the model as a function of current density. At the cathode the methanol that crosses the membrane is oxidized in a corrosion reaction. The leakage current cannot be used to do work. Expressing the methanol crossover, as in Fig. 6, in terms of the leakage current gives a more tangible understanding of the loss in efficiency due to methanol crossover. The leakage current can be reduced by running the cell at low methanol concentrations and high current densities. Thus to reduce crossover running at lower concentrations of methanol may be advantageous. The leakage currents calculated in this paper are similar to those calculated by Wang and Wang [12]. It should be noted that the leakage current goes to zero at the limiting current value for all concentrations. This provides a check that our transport equations are giving a physically meaningful concentration profile.

## CONCLUSIONS

A semi- analytical, 1-D, isothermal model of a DMFC has been developed. Using reasonable transport and kinetic parameters the model fits well to experimental polarization data. The model allows prediction of concentration profiles in the anode and membrane as well as estimating methanol crossover. The solution time is less than 1 minute.

## ACKNOWLEDGMENT

This work was carried out under Agreement No. DAAB07-03-3-K416 with the US Army Communications-Electronics Command (CECOM) for Hybrid Advanced Power Sources with guidance from the RDECOM / CERDEC Fuel Cell Technology Team at Fort Belvoir, VA.

# NOMENCLATURE

| | |
|---|---|
| $a$ | specific surface area of the anode, $cm^{-1}$ |
| $c_b$ | bulk concentration of methanol in the flow channel, $mol/cm^3$ |
| $c_I$ | concentration of methanol at the ABL/ACL interface, $mol/cm^3$ |
| $c_{II}$ | concentration of methanol at the ACL/membrane interface, $mol/cm^3$ |
| $c_{MeOH}$ | concentration of methanol, $mol/cm^3$ |
| $c_{O_2}$ | concentration of oxygen, $mol/cm^3$ |
| $c^G$ | total concentration in the ABL, $mol/cm^3$ |
| $D_A$ | effective diffusion coefficient of methanol in the ACL, $cm^2/s$ |
| $D_B$ | effective diffusion coefficient of methanol in the ABL, $cm^2/s$ |
| $D_M$ | effective diffusion coefficient of methanol in the membrane, $cm^2/s$ |
| $F$ | Faraday's constant, 96,487 C/equiv |
| $I_{Cell}$ | cell current density, $A/cm^2$ |
| $I_{leak}$ | leakage current density due to methanol crossover, $A/cm^2$ |
| $I_{0,ref}^{MeOH}$ | exchange current density of methanol, $A/cm^2$ |
| $I_{0,ref}^{O_2}$ | exchange current density of oxygen, $A/cm^2$ |
| $j$ | volumetric current density, $A/cm^3$ |
| $k$ | constant in the rate expression (Eq. 21), dimensionless |
| $M_{MeOH}$ | molecular weight of methanol, g/mol |
| $N_{z,MeOH}$ | z component of methanol molar flux, mol/(cm² s) |
| $R$ | gas constant, 8.314 J/(mol K) |
| $r_{MeOH}$ | rate of consumption of methanol by homogeneous reaction, g/ (cm³ s) |
| $T$ | temperature, K |
| $U^{MeOH}$ | thermodynamic equilibrium potential of methanol oxidation, V |
| $U^{O_2}$ | thermodynamic equilibrium potential of oxygen oxidation, V |
| $V_{Cell}$ | cell voltage, V |
| $x_{MeOH}$ | mole fraction of methanol, mol/mol |
| $z$ | coordinate direction normal to the anode, cm |

Greek

| | |
|---|---|
| $\alpha_A$ | anodic transfer coefficient |
| $\alpha_C$ | cathodic transfer coefficient |
| $\delta_A$ | ACL thickness, cm |
| $\delta_B$ | ABL thickness, cm |
| $\delta_M$ | membrane thickness, cm |
| $\eta_A$ | anode overpotential, V |
| $\eta_C$ | mix overpotential at the cathode, V |
| $\kappa$ | ionic conductivity of the membrane, S/cm |
| $\lambda$ | constant in the rate expression (Eq. 21), mol/cm$^3$ |
| $\xi_{MeOH}$ | electro-osmotic drag coefficient of methanol |

Subscripts

| | |
|---|---|
| A | ACL |
| B | ABL |
| b | bulk |
| Cell | cell |
| I | ABL/ACL interface |
| II | ACL/membrane interface |
| III | membrane/cathode layer interface |
| M | membrane |
| MeOH | methanol |
| O$_2$ | oxygen |
| z | z-direction |

Superscripts

| | |
|---|---|
| A | ACL |
| B | ABL |
| M | membrane |
| MeOH | methanol |
| $O_2$ | oxygen |

# REFERENCES

1. Cruickshank, J., and Scott, K., (1998), "The Degree and Effect of Methanol Crossover in the Direct Methanol Fuel Cell," *J Power Sources*, **70** (1), pp. 40-47.

2. Scott, K., Taama, W. M., Argyropoulos, P., and Sundmacher, K., (1999), "The Impact of Mass Transport and Methanol Crossover on the Direct Methanol Fuel Cell," *J Power Sources*, **83** (1-2), pp. 204-216.

3. Ren, X., Springer, T. E., and Gottesfeld, S., (2000), "Water and Methanol Uptakes in Nafion Membranes and Membrane Effects on Direct Methanol Cell Performance," *J Electrochem Soc*, **147** (1), pp. 92-98.

4. Dohle, H., Divisek, J., Merggel, J., Oetjen, H. F., Zingler, C., and Stolten, D., (2002), "Recent Developments of the Measurement of the Methanol Permeation in a Direct Methanol Fuel Cell," *J Power Sources*, **105** (2), pp. 274-282.

5. Carrette, L., Friedrich, K. A., and Stimming, U., (2001), "Fuel Cells - Fundamentals and Applications," *Fuel Cells*, **1** (1), pp. 5-39.

6. Gasteiger, H. A., Markovic, N. M., and Ross, P. N., (1995), "$H_2$ and CO Electrooxidation on Well-Characterized Pt, Ru, and Pt-Ru .1. Rotating-Disk Electrode Studies of the Pure Gases Including Temperature Effects," *J Phys Chem-Us*, **99** (20), pp. 8290-8301.

7. Iwasita, T., (2002), "Electrocatalysis of Methanol Oxidation," *Electrochim Acta*, **47** (22-23), pp. 3663-3674.

8. Desai, S., and Neurock, M., (2003), "A First Principles Analysis of CO Oxidation over Pt and $Pt_{66.7\%}Ru_{33.3\%}$ (111) Surfaces," *Electrochim Acta*, **48** (25-26), pp. 3759-3773.

9. Meyers, J. P., and Newman, J., (2002), "Simulation of the Direct Methanol Fuel Cell - Ii. Modeling and Data Analysis of Transport and Kinetic Phenomena," *J Electrochem Soc*, **149** (6), pp. A718-A728.

10. Baxter, S. F., Battaglia, V. S., and White, R. E., (1999), "Methanol Fuel Cell Model: Anode," *J Electrochem Soc*, **146** (2), pp. 437-447.

11. Kulikovsky, A. A., (2003), "Analytical Model of the Anode Side of DMFC: The Effect of Non-Tafel Kinetics on Cell Performance," *Electrochem Commun*, **5** (7), pp. 530-538.

12. Wang, Z. H., and Wang, C. Y., (2003), "Mathematical Modeling of Liquid-Feed Direct Methanol Fuel Cells," *J Electrochem Soc*, **150** (4), pp. A508-A519.

13. Nordlund, J., and Lindbergh, G., (2002), "A Model for the Porous Direct Methanol Fuel Cells Anode," *J Electrochem Soc*, **149** (9), pp. A1107-A1113.

14. Wilson, M. S., 1993, U. S. Patent 5,211,984.

15. Slattery, J. C., (1999), *Advanced Transport Phenomena*, Cambridge University Press, Cambridge, MA.

16. Scott, K., Taama, W., and Cruickshank, J., (1997), "Performance and Modelling of a Direct Methanol Solid Polymer Electrolyte Fuel Cell," *J Power Sources*, **65** (1-2), pp. 159-171.

17. Parthasarathy, A., Srinivasan, S., Appleby, A. J., and Martin, C. R., (1992), "Temperature Dependence of the Electrode Kinetics of Oxygen Reduction at the

Platinum/Nafion Interface - a Microelectrode Investigation," *J Electrochem Soc*, **139** (9), pp. 2530-2537.

18. Ren, X. M., Springer, T. E., Zawodzinski, T. A., and Gottesfeld, S., (2000), "Methanol Transport through Nafion Membranes - Electro-Osmotic Drag Effects on Potential Step Measurements," *J Electrochem Soc*, **147** (2), pp. 466-474.

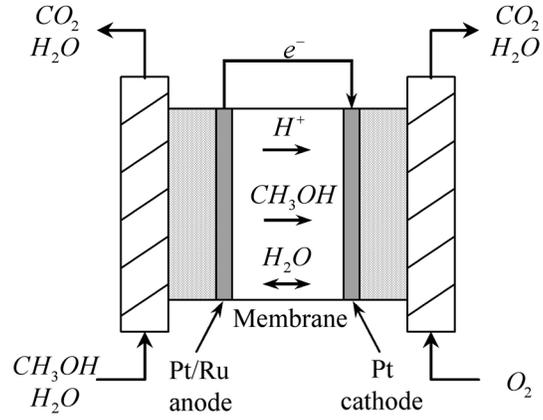

Anode: $CH_3OH + H_2O \rightarrow CO_2 + 6H^+ + 6e^-$

Cathode: $\frac{3}{2}O_2 + 6H^+ + 6e^- \rightarrow 3H_2O$

Cathode corrosion reaction: $CH_3OH + H_2O \rightarrow CO_2 + 6H^+ + 6e^-$

Overall: $\frac{3}{2}O_2 + CH_3OH \rightarrow CO_2 + 2H_2O$

**Figure 1.** Schematic of a DMFC.

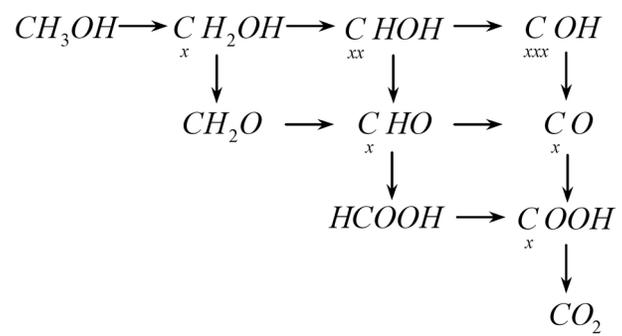

**Figure 2.** Reaction pathways of methanol oxidation [5].

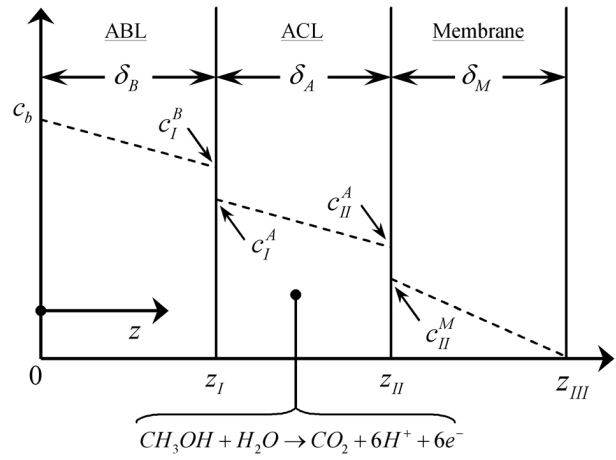

**Figure 3.** Schematic of the DMFC layers considered in the model.

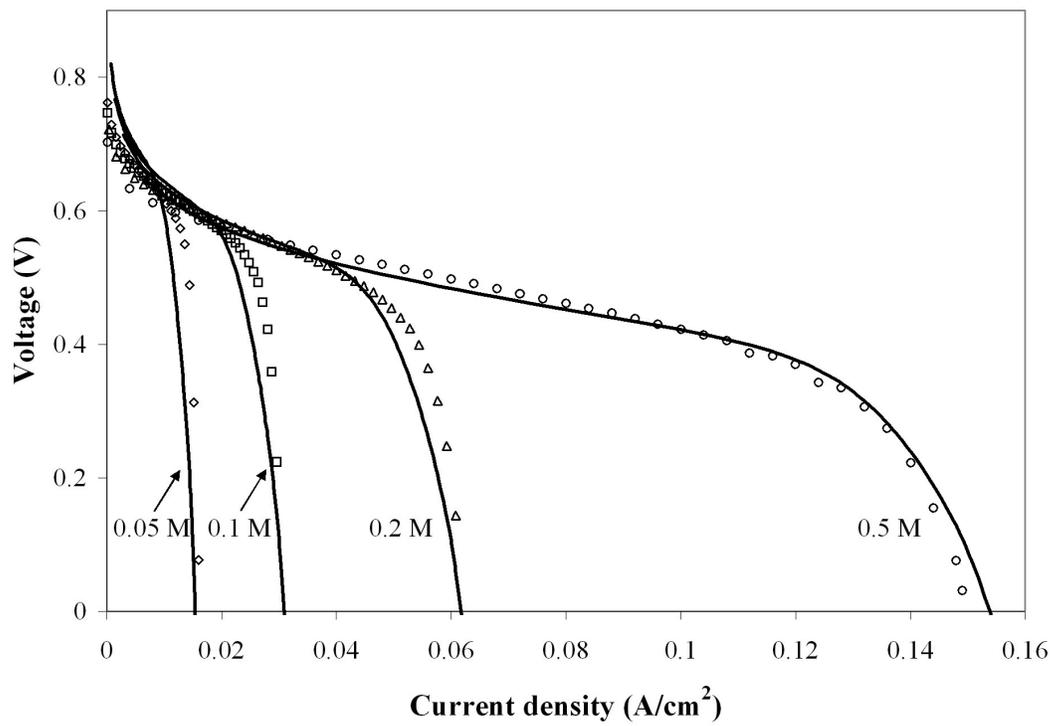

**Figure 4.** Model predictions for different methanol concentrations.

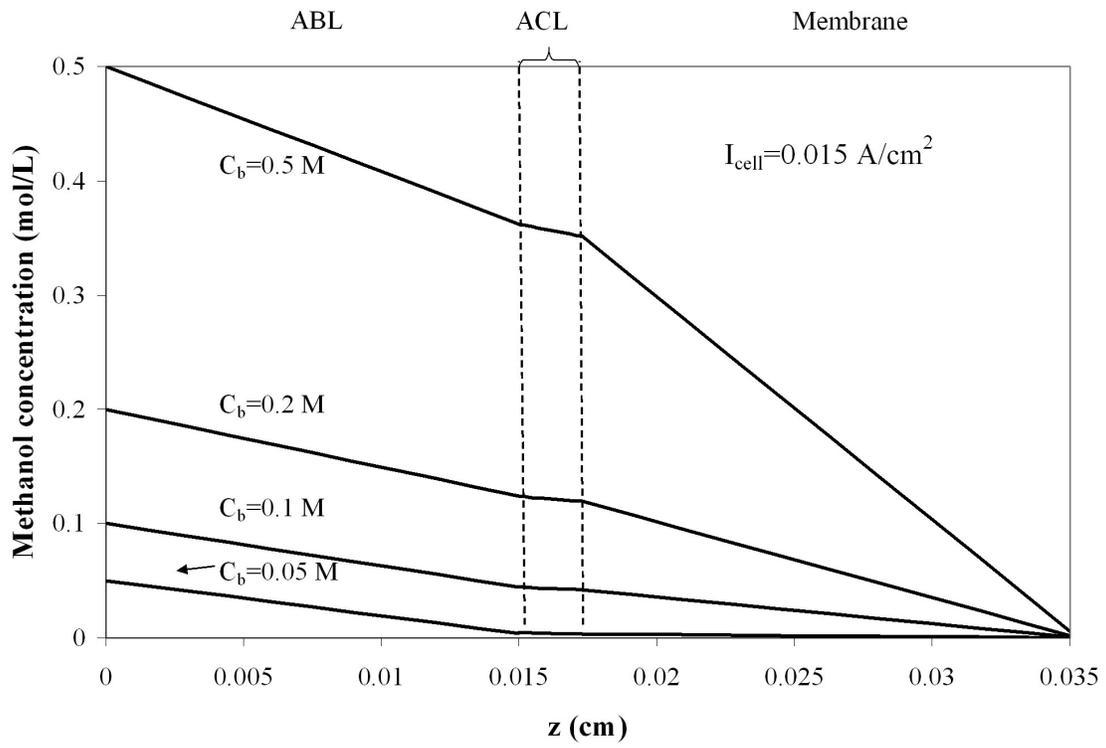

**Figure 5.** Concentrations profiles for different methanol bulk concentrations.

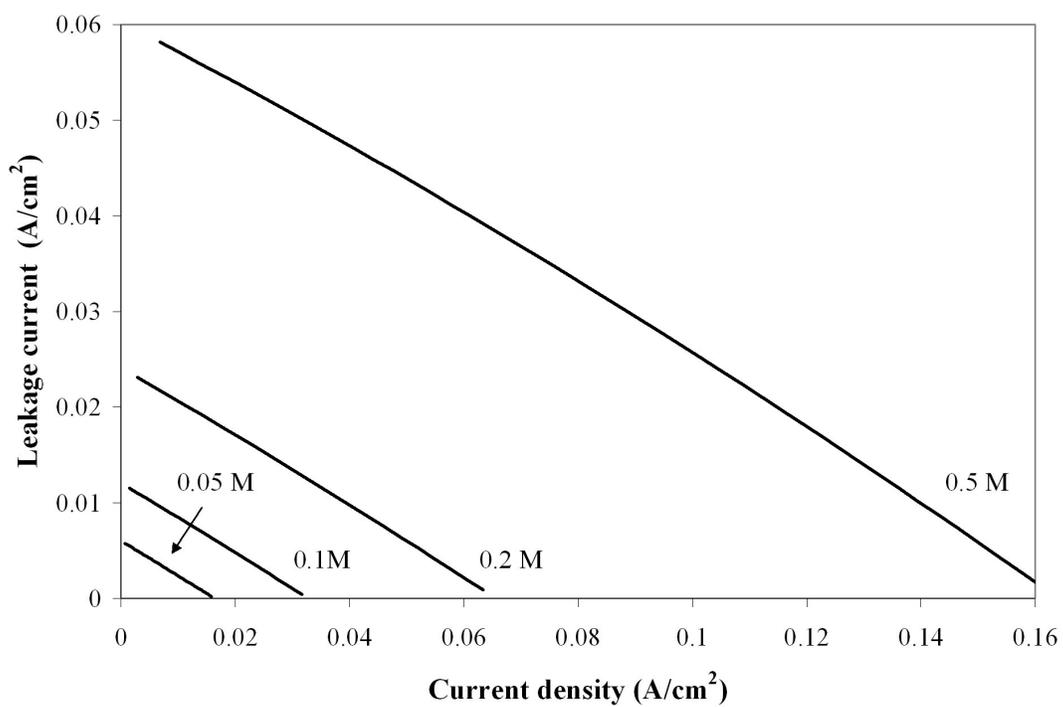

**Figure 6.** Methanol crossover for different methanol bulk concentrations.

**Table 1. Parameter Values.**

| Parameter | Value | Ref. |
|---|---|---|
| $a$ | 1000 cm$^2$ | Assumed |
| $D_A$ | $2.8 \times 10^{-5} e^{2436\left(\frac{1}{353} - \frac{1}{T}\right)} \frac{\text{cm}^2}{\text{s}}$ | Scott et al. [16] |
| $D_B$ | $8.7 \times 10^{-6}$ cm$^2$/s | Assumed |
| $D_M$ | $4.9 \times 10^{-6} e^{2436\left(\frac{1}{333} - \frac{1}{T}\right)} \frac{\text{cm}^2}{\text{s}}$ | Scott et al. [16] |
| $I_{0,ref}^{MeOH}$ | $9.425 \times 10^{-3} e^{\frac{35570}{R}\left(\frac{1}{353} - \frac{1}{T}\right)} \frac{\text{A}}{\text{cm}^2}$ | Wang and Wang [12] |
| $I_{0,ref}^{O_2}$ | $4.222 \times 10^{-3} e^{\frac{73200}{R}\left(\frac{1}{353} - \frac{1}{T}\right)} \frac{\text{A}}{\text{cm}^2}$ | Parthasarathy et al. [17] |
| $K_I$ | 0.8 | Baxter et al. [10] |
| $K_{II}$ | 0.8 | Baxter et al. [10] |
| $k$ | $7.5 \times 10^{-4}$ | Assumed |
| $T$ | 343.15 K | ----- |
| $U^{MeOH}$ | 0.03 V | Wang and Wang [12] |
| $U^{O_2}$ | 1.24 V | Wang and Wang [12] |
| $\alpha_A$ | 0.52 | Assumed |
| $\alpha_C$ | 1.55 | Assumed |
| $\delta_A$ | 0.0023 cm | ----- |
| $\delta_B$ | 0.015 cm | ----- |
| $\delta_M$ | 0.018 cm | ----- |
| $\kappa$ | 0.036 S/cm | Assumed |
| $\lambda$ | $2.8 \times 10^{-9}$ mol/cm$^3$ | Assumed |
| $\xi_{MeOH}$ | $2.5 x_{MeOH}$ | Ren et al. [18] |

# Errata: "Mathematical Model of a Direct Methanol Fuel Cell"

B. L. García, V. A. Sethuraman, J. W. Weidner,[1] R. E. White, and R. Dougal

The authors inform that there are three errors in the article.

(1) The following correction applies to Eq. (30). Equation (30) contains a parameter $n$ that was not defined. The parameter $n$ is the number of electron transferred in the methanol oxidation reaction. Hence $n = 6$. Substituting this value into Eq. (30) give the corrected equation as

$$c_{II}^{A} = \frac{\delta_M \left( D_A D_B c_b - \delta_A D_B K_I (1 + 12\xi_{MeOH}) \frac{I_{Cell}}{12F} - \delta_B D_A (1 + 6\xi_{MeOH}) \frac{I_{Cell}}{6F} \right)}{D_B K_I (\delta_A D_M K_{II} + \delta_M D_A) + \delta_B D_A D_M K_{II}}$$

(2) The following corrections apply to Table 1

$$\lambda = 5.5 \times 10^{-9} \, \text{mol/cm}^3$$
$$D_B = 1 \times 10^{-5} \, \text{cm}^2/\text{s}$$
$$K_I = 1.25$$

(3) The figure below corresponds to the corrected Fig. 5

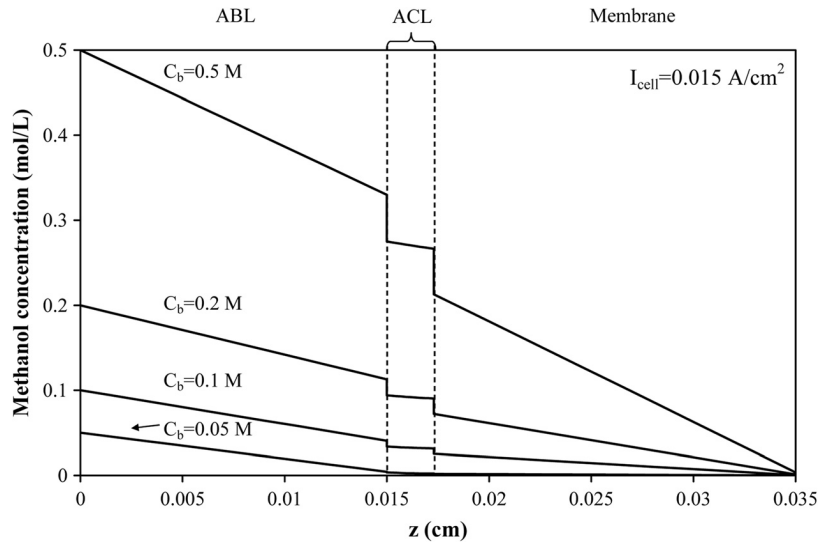

**Fig. 5** Concentrations profiles for different methanol bulk concentrations

---
[1]Corresponding author: weidner@engr.sc.edu.